\documentclass[
    a4paper,
    jou,
    american,
    floatsintext
]{apa6}
\pdfoutput=1

\usepackage{amsmath}
\usepackage{amssymb}
\usepackage{setspace}
\usepackage{booktabs}
\usepackage{placeins}
\usepackage{titletoc}
\usepackage{lineno}
\usepackage{wrapfig}
\usepackage{multirow}
\usepackage[american]{babel}
\usepackage{epstopdf}
\usepackage{csquotes}
\usepackage{color}
\usepackage{apacite}
\usepackage{framed}
\usepackage{subcaption} % For two figures in appendix
\usepackage{caption,graphicx,newfloat}
\usepackage{xr}
\usepackage{url}
\usepackage{multicol}

\title{\textbf{Group Decisions Based on Confidence Weighted Majority Voting}}
\shorttitle{Group Decisions Based on CWMV}

\fourauthors{Sascha Meyen}{Dorothee M. B. Sigg}{Ulrike von Luxburg}{Volker H. Franz}

\fiveaffiliations{Experimental Cognitive Science, University of Tübingen}{Experimental Cognitive Science, University of Tübingen}{Theory of Machine Learning, University of Tübingen \\ Max Planck Institute for Intelligent Systems, Tübingen}{Experimental Cognitive Science, University of Tübingen}

\rightheader{Group Decisions Based on Confidence Weighted Majority Voting}
\leftheader{Meyen, Sigg, von Luxburg, \& Franz}

\AtBeginDocument{%
}

\abstract{
\singlespacing
\noindent\textbf{Background:} It has repeatedly been reported that, when making decisions under uncertainty, groups outperform individuals. Real groups are often replaced by simulated groups: Instead of performing an actual group discussion, individual responses are aggregated by a numerical computation. While studies have typically used unweighted majority voting (MV) for this aggregation, the theoretically optimal method is confidence weighted majority voting (CWMV)---if independent {\color{black}and accurate} confidence ratings from the individual group members are available. To determine which simulations (MV vs. CWMV) reflect real group processes better, we applied formal cognitive modeling and compared simulated group responses to real group responses. \newline

\noindent\textbf{Results:} Simulated group decisions based on CWMV matched the accuracy of real group decisions, while simulated group decisions based on MV showed lower accuracy. CWMV predicted the confidence that groups put into their group decisions well. However, real groups {\color{black}treated} individual votes to some extent more equally weighted than suggested by CWMV. Additionally, real groups tend to put lower confidence into their decisions compared to CWMV simulations. \newline

\noindent\textbf{Conclusion:} Our results highlight the importance of taking individual confidences into account when simulating group decisions: We found that real groups can aggregate individual confidences {\color{black}in a way that matches statistical aggregations} given by CWMV to some extent. This implies that research using simulated group decisions should use CWMV instead of MV as a benchmark to compare real groups to. \newline

}


\begin{document}
\maketitle

%%%%%%%%%%%%%%%%%%%%%%%%% start of article main body
\singlespacing
\section*{Significance Statement}

The question of how a group determines an overall group decision from the individual votes of its group members is pervasive and likely as old as mankind. It is at the basis of democratic voting rules and is also prevalent with new urgency in the age of the Internet, where often many individual votes, or ratings, are available that one wants to combine to an optimal overall group decision---without there being the possibility of real group discussions.
From a theoretical point of view, the situation is clear: Individual confidences should be taken into account and confidence weighted majority voting (CWMV) is the statistically optimal aggregation procedure (under quite general assumptions). 
However, in research on group decisions, CWMV is not routinely used for comparison to real group performances, but instead the simpler majority vote (MV) that ignores the individual confidences. Therefore, it is currently not clear whether real groups weigh individual votes in the same way CWMV does. Real groups may be limited in their capacity to take individual confidence ratings into consideration or may rely on different strategies.
We compared real group decision to simulated group decisions based on the CWMV and MV procedures. We found that real groups weigh individual confidences in a way that can be well described by CWMV. These results suggest that basic research as well as online-based aggregation of individual votes or ratings could benefit from using CWMV instead of MV. 

%%%%%%%%%%%%%%%%
%% Background %%
%%
\section*{Background}

Under uncertainty, groups make more accurate decisions than individuals \cite{koriat2015two,mannes2014wisdom}: Medical students achieve more accurate diagnoses in groups than individually \cite{hautz2015}; medical diagnoses improve when groups of independent doctors are involved \cite{kurvers2016boosting,wolf2015collective}; groups of students make more accurate judgments about criminal cases than individuals \cite{vandijk2014}; groups detect lies more accurately than individuals \cite{klein2015}; groups achieve higher IQ scores than individuals (referred to as wisdom of the crowd, \citeNP{bachrach2012crowd,vercammen2019collective,kosinski2012crowd}) etc. Exceptions occur when group members have widely different levels of competence \cite{galesic2018smaller,puncochar2004confidence,vandijk2014}. Nevertheless, groups generally outperform individuals.

Although some of the above mentioned studies also used real groups \cite{hautz2015,klein2015,vandijk2014}, all of these studies simulated group decisions: Individuals gave responses that were then statistically aggregated into one simulated group response without a real group discussion occurring. A crucial aspect is therefore the choice of aggregation method that is used to simulate group decisions. One frequently used method is majority voting (MV; \citeNP{hastie2005}; and see for example \citeNP{klein2015,vandijk2014,kosinski2012crowd,kurvers2016boosting,sorkin2001signal}). 

In MV, the most frequent individual decision (vote) is taken as the simulated group decision. By design, MV weighs all individual responses equally. Note, however, that real groups typically perform better than simulated groups using MV \cite{bahrami2010,birnbaum2015testing,klein2015,sniezek1989accuracy}. This shows that MV cannot capture all the processes that are at work in real group decisions.

In particular, MV overlooks that individuals can estimate how accurate their own decisions are in many situations \cite{brenner1996,fleming2012,griffin1992,martins2006probability,zehetleitner2013being,regenwetter2014qtest} even though there are also situations in which they cannot \cite{klein2015,koriat2012two,koriat2017can,litvinova2019experts}. When reliable confidence estimates are available, they can influence real group discussions: It is plausible that individuals share their sense of confidence during group interactions \cite{bang2014does} such that votes from confident individuals are weighted more than those of less confident individuals.

There are methods that have taken confidence ratings from individuals into account. One of the most prominent is the maximum-confidence slating algorithm by \citeA{koriat2012self,koriat2012two}. In this algorithm, the most confident individual decides the vote. Another approach for dealing with multiple confidence ratings is to not only consider the most confident individual but a small subgroup of the top most confident individuals \cite{mannes2014wisdom}, or to average all confidences \cite{litvinova2020wisdom}. However, these methods to simulate group decisions do not strictly follow the mathematically optimal way to aggregate confidences. 

The theoretically optimal method to aggregate individual confidences is confidence weighted majority voting (CWMV; \citeNP{grofman1983,NitzanParoush1982})—assuming that individuals can accurately assess confidences in their independently formed decisions. CWMV aggregates these independent responses (votes and confidences) in the mathematically optimal way by giving more weight to reliable than unreliable votes. Thus, statistically aggregating individual responses into a simulated group decision using CWMV rather than MV may reflect real groups better and provide a more appropriate benchmark.

Do real, interacting groups weigh individual confidences in a way that is reflected by simulating a group discussion using CWMV? It is not clear whether real group decisions are adequately represented by CWMV, since CWMV is only sporadically applied in current research. \citeA{bahrami2010} found that group performance of dyads is well predicted by CWMV. \citeA{hautz2015} found that real dyads performed better than CWMV, which predicts the group response of a dyad to be that of the most confident member. CWMV is also discussed in animals from an evolutionary perspective \cite{marshall2017}. However, to our knowledge, no study has yet considered groups with more than two members compar{\color{black}ing} decisions {\color{black}from} real group discussions versus simulated decisions using CWMV on a trial-by-trial basis.

In our experiment, we investigated whether CWMV simulations can predict real group decision of triads (groups of three). We compared simulated group decisions to real group decisions on a trial-by-trial basis. Our groups consisted of three individuals because we wanted to investigate whether real groups weigh confidences in a way that is adequately reflected by CWMV. In contrast, using only dyads, CWMV simulates the group decisions to be the vote of the more confident individual (similar to maximum-confidence slating) and CWMV can only contribute by predicting a dyad's combined confidence based on the individual responses. But triads can display qualitatively different behavior than dyads: While it is sometimes the case that the most confident individual determines the group decision in triads, triads also allow for the possibility that the most confident individual is overruled by the two other group members when they are sufficiently confident in the alternative choice. Thus, we want to clarify whether real groups of three weigh individual votes in a way that can be characterized by CWMV.

Before describing our experiment, we will give a more formal description of the simulation methods MV and CWMV. We will present a formal cognitive model (e.g., see \citeNP{forstmann2011reciprocal}) that allows us to measure in how far real groups deviate from CWMV simulations.

\subsection*{Majority voting (MV) versus confidence weighted majority voting (CWMV)}

CWMV assumes that multiple individuals report independent decisions (votes) as well as confidence ratings. These confidence ratings indicate how reliable individual decisions are. CWMV weighs the decisions by the confidence ratings in a theoretically optimal way to simulate a group decision \cite{grofman1983,NitzanParoush1982}. This section shortly introduces the basic mathematical notation, first of MV and then of CWMV.

Let a group consist of $n$ individuals. The task is to decide between multiple (usually two) options from which exactly one is correct. For example, consider $n = 3$ students trying to determine whether a suspect of a criminal case is guilty or not (cf. \citeNP{vandijk2014}). First, each individual forms a decision $y_i$ which is either $+1$ (not guilty) or $-1$ (guilty). Second, in a real, interactive group discussion, the individual group members reach a common decision $y_g$. 

The real-world group decision $y_g$ can be simulated by statistically aggregating the independently formed individual responses. MV simulates the group decision to be that of the majority of individuals, $y^{MV}_g = \text{sign}(\sum_{i=1}^n y_i)$. MV (as well as CWMV) assumes that individual responses are independent from each other given the ground truth, that is, individuals must form their {\color{black}decision} only based on material that is not systematically shared between members. To illustrate a violation of this assumption, consider as another example a group of radiologists forming their individual diagnoses based on one and the same x-ray. They will not come to fully independent conclusions about the true state of the patient's condition because their opinions will be commonly influenced by the quality of the x-ray. In the worst case, multiple individual responses are fully dependent offering no more information than one single response. In our experiment, independence will be ensured by design in order to study CWMV—even though many real world situations will not allow for such a controlled environment.

When individuals report confidence ratings, $c_i$, MV can be improved upon by using CWMV instead. These confidence ratings are assumed to be in the form of estimates for the probability of their decision being correct, $c_i = P(y_i \text{ is correct})$. In some situations, individuals can make such estimates \cite{griffin1992,martins2006probability,regenwetter2014qtest,koriat2012self} and, under specific circumstances, assessing confidences is essentially the same as estimating the relative frequency of being correct \cite{brenner1996,pouget2016confidence}. CWMV transforms these confidences into optimal weights, which are the logarithmic odds (log odds), $w_i = \log(c_i/(1-c_i))$. See \citeA{NitzanParoush1982} as well as \citeA{shapley1984optimizing}, and find an intuitive account for using logarithmic odds as weights at the end of this section. Using these weights, CWMV simulates the group decision by 
\begin{linenomath*}
    \begin{equation}
        y^{\text{CWMV}}_g = \text{sign}\left(\sum_{i=1}^n w_i y_i\right) \text{ .} \label{eqn:ycwmv}
    \end{equation}
\end{linenomath*}
Similar to the individual confidence ratings, real groups can also report how confident they are in their group decision $c_g$. CWMV can simulate these group confidences based on the individual confidences by 
\begin{linenomath*}
    \begin{equation}
        c^{\text{CWMV}}_g = \frac{1}{1+\exp(-|\sum_{i=1}^n w_i y_i|)}\text{ .} \label{eqn:ccwmv}
    \end{equation}
\end{linenomath*}

To illustrate the computation of CWMV, consider again the three students deciding whether a suspect is guilty. Say, Student~1 votes for the suspect being innocent, $y_1 = +1$, but Students 2 and 3 believe the suspect to be guilty, $y_2 = -1$ and $y_3 = -1$. Aggregating these decisions using MV determines the simulated group decision to be guilty, $y^{MV}_g = \text{sign}((+1) + (-1) + (-1)) = -1$. Additionally, Student~1 reports being quite confident in their vote such that the probability of their judgment being correct is 76\%, $c_1 = .76$. In contrast, Students~2 and 3 are very unsure with a confidence of only 51\%, $c_2 = c_3 = .51$. Using CWMV to integrate these individual responses into a simulated group decision, the individual confidences are first transformed into weights with Student 1 having a higher confidence and, thus, a larger weight: $w_1 = \log(.76/.24) = 1.15$ versus $w_2 = w_3 = \log(.51/.49) = 0.04$. Then, CWMV leads to a different simulated group decision than MV finding the suspect not guilty, $y^{\text{CWMV}}_g = \text{sign}\left((+1.15) + (-0.04) + (-0.04)\right) = \text{sign}(+1.07) = +1$. Moreover, CWMV simulates the group's confidence in their verdict to be 75\%, $c^{\text{CWMV}}_g = 1/[1+\exp(- |(+1.07)|)] = .75$. That is, the confident response from Student~1 is only slightly attenuated by the unconfident, opposing responses from Students~2 and 3 as might be realistic in a real group discussion. This example corresponds numerically to Scenario~II from our experiment, which we use to study in how far real groups are better represented by MV or CWMV simulations, see Table~\ref{tab:scenarios}.

\begin{table*}[!ht]
\caption{\textbf{Ideal decisions and confidences.} In each trial, we applied one out of four scenarios (I--IV) which is defined by three stimulus sequences (A, B and C). Each of the three participants from a group viewed one stimulus sequence. Each individual stimulus sequence entails an ideal decision $y^*_i$ and ideal confidence $c^*_i$ that can be derived from probability computations. The ideal individual responses from each scenario determine the groups' ideal decision $y^*_g$ and confidence $c^*_g$ (see Methods section for an example calculation corresponding to Scenario~II).}
\label{tab:scenarios}
\centering
      \begin{tabular}{lllllllll}

        \toprule
        Scenario    & \multicolumn{6}{c}{Individual}                                        & \multicolumn{2}{c}{Group} \\ 
                     \cmidrule(lr){2-7} \cmidrule(lr){8-9}
                    & \multicolumn{2}{c}{A} & \multicolumn{2}{c}{B} & \multicolumn{2}{c}{C} &           & \\ 
                    \cmidrule(lr){2-3} \cmidrule(lr){4-5} \cmidrule(lr){6-7}
            & $y^*_1$     & $c^*_1$     & $y^*_2$     & $c^*_2$     & $y^*_3$     & $c^*_3$     & $y^*_g$     & $c^*_g$ \\[-.75em] \\ \hline
        \\[-.75em]
        I           & fair coin      & 87\%      & fair coin      & 70\%      & fair coin      & 62\%      & fair coin      & 96\%  \\
        \\[-.75em]
        II          & biased coin      & 76\%      & fair coin      & 51\%      & fair coin      & 51\%      & biased coin      & 75\%  \\
        \\[-.75em]
        III         & biased coin      & 88\%      & biased coin      & 54\%      & fair coin      & 81\%      & biased coin      & 66\%  \\
        \\[-.75em]
        IV          & fair coin      & 81\%      & biased coin      & 58\%      & biased coin      & 72\%      & fair coin      & 54\%  \\
        \bottomrule
      \end{tabular}
\end{table*}

A technical note: The weights in CWMV are log odds because the logarithm is used as a convenient trick to transform a multiplication into a weighted sum. When computing the probabilities of a suspect being guilty or not, basic Probability Theory gives that odds ($o_i = c_i/(1-c_i)$) can be multiplied. In our example, the odds of the suspect being innocent are $o_1 = .76/.24$ and $o_2 = o_3 = .51/.49$. Multiplying these odds results in group odds, $o_g = o_1\times o_2^{-1} \times o_3^{-1} = 3$ ($o_2$ and $o_3$ are inverted because Student 2 and 3 vote for guilty). Observe, that the group odds are indeed equivalent to the 75\% group confidence computed by CWMV from above, $c_g/(1-c_g) = .75/.25 = 3$. By applying the laws of logarithm, multiplication of these odds is transformed into a sum of the log odds: $\log(o_1\times o_2^{-1} \times o_3^{-1}) = (+\log(o_1)) + (-\log(o_2)) + (-\log(o_3))$, which allows to derive Equations~\ref{eqn:ycwmv} and \ref{eqn:ccwmv}. Note further that, when an individual is absolutely certain in their decision ($c_i = 0$ or $c_i = 1$), the odds $o_i$ and weights $w_i$ are undefined. In this case, by convention, the simulated group is set to be absolutely certain as well ($c_g = 0$ or $c_g = 1$). But if two participants came to opposite decisions and were both absolutely certain, by convention, their two responses would be discarded and the third individual's vote would decide (this situation did not occur in our experiment).

Given this formal framework of CWMV, the purpose of this study is to investigate how well individual responses ($y_i$ and $c_i$) aggregated into simulated group responses ($y^{\text{CWMV}}_g$ and $c^{\text{CWMV}}_g$) represent the real group responses from actual group discussions ($y_g$ and $c_g$) on a trial-by-trial basis. We will modify Equations~\ref{eqn:ycwmv} and \ref{eqn:ccwmv} using formal cognitive modeling in order to characterize how real groups deviate from these CWMV simulations.

\section*{Methods}

\subsection*{Participants}

A total of 21 participants (11 female, mean age = 21.4, range = 19 - 26) completed the experiment in seven groups of three. All were students who received either course credit for $30$~min of participation or payment ($4$ EUR, equivalent to $4.5$ USD). All participants had normal or corrected-to-normal vision and provided written informed consent prior to participation.

\subsection*{Stimuli \& procedure}

\begin{figure*}[!ht]
    \includegraphics[page=1,width=.95\textwidth]{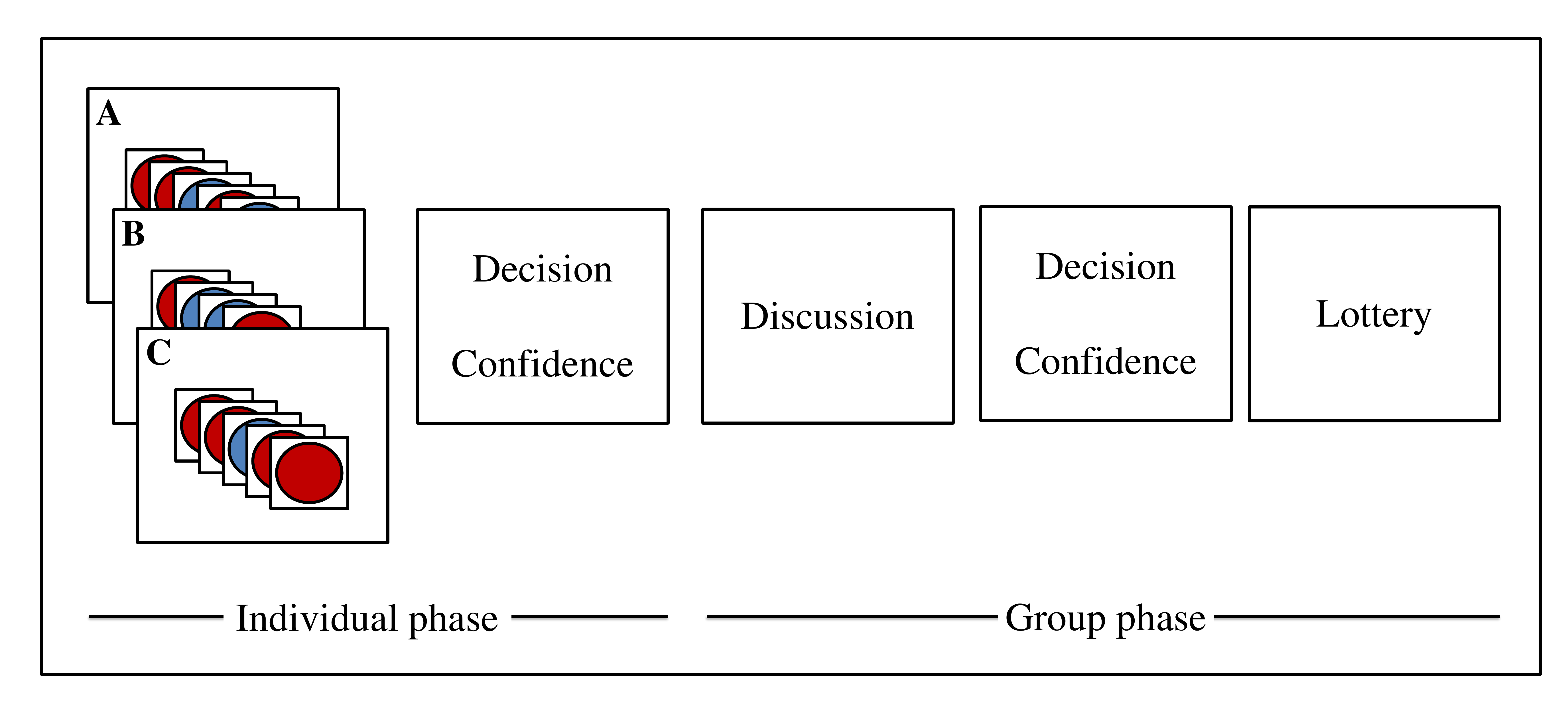}
    \caption{\textbf{Individual and group phase in each trial.} In the individual phase, each participant viewed different stimulus sequences consisting of 11 to 13 disks. Based on these sequences, individuals decided whether their sequence has more likely been produced by a fair coin (50\% red, 50\% blue) or a biased coin (60\% red, 40\% blue). Based on the ambiguity of the sequence, individuals reported a confidence in their own decision. In the group phase, participants combined their evidence into one group decision and confidence. In each trial, participants were incentivized for accurately judging their real group confidence using the Matching Probability method by \protect\citeA{massoni2014}.}
      \label{fig:procedure}
\end{figure*}

We adopted a procedure that has been established by \citeA{griffin1992} and extended it to a group setting. The experiment consisted of three practice trials followed by 12 experimental trials. Each trial consisted of an individual phase and a group phase, see Figure~\ref{fig:procedure}. Participants viewed rapid stimulus sequences consisting of 11 to 13 red and blue disks. Their task was to guess whether the stimulus sequence was generated by a fair coin (producing in expectation 50\% red and 50\% blue disks) or a biased coin (producing 60\% red and 40\% blue disks). Participants were instructed that both, the fair and the biased coin, are a priori equally likely. \citeA{griffin1992} showed that, in this task, participants' individual confidence ratings are well calibrated.

Participants viewed different stimulus sequences simultaneously at individual laptops. Their viewing distance to the screen was approximately $60$~cm. Each disk was presented for $100$~ms with a diameter of $2.2$~cm corresponding to a viewing angle of $2.1^{\circ}$. Disks were intermitted by a $100$~ms blank interval creating the impression of a rapid stream. This presentation prevented participants from performing explicit mathematical calculations so that they could only obtain an intuitive sense of confidence.

Depending on which coin better matched the stimulus sequence, participants made a decision for either the fair or the biased coin. Some stimulus sequences were more ambiguous than others providing different levels of confidence that participants reported on a visual analog scale from $50\%$ (\enquote{I am completely unsure. The other option is equally likely.}) to $100\%$ (\enquote{I am completely sure. My decision is definitively correct.}). Participants were instructed to report their subjective probability with which they believed their decision to be correct. In contrast to verbal scales, where participants give responses such as ``somewhat likely'' or ``almost certain'', this numeric scale is necessary because numeric values (here, $c_i$) are required to simulate group decisions (see Equation~\ref{eqn:ycwmv}). However, it is noteworthy that a numerical scale can prompt participants to engage in formal thinking, which they would otherwise have not \cite{windschitl1996measuring}.

The presented stimulus sequences determined ideal individual responses, which reflect posterior probabilities that can be computed using Probability Theory. Table~\ref{tab:scenarios} shows which responses the stimulus sequences would produce if participants were ideal observers. For example, assume that a participant saw the disk sequence red, red, blue, red and red. A fair coin would have produced such a sequence with a likelihood of $p_\text{fair} = 0.5^5 = 3\%$ and the biased coin with $p_\text{biased} = 0.6^4\cdot 0.4^1 = 5\%$. Because the biased coin was more likely to produce this stimulus sequence, the ideal decision is for the biased coin denoted by $y^*_i = +1$ (the asterisk denotes ideal values). The ideal confidence was $c^*_i = p_\text{biased}/(p_\text{fair} + p_\text{biased}) = 5\%/(3\%+5\%) = 62\%$. 

Note that scheduling individual reports before a group discussion (as in our experiment) improves group performance and prevents contamination of individual reports by the group decision \cite{sniezek1990revision}. That is, individual reports remain independent because participants interacted only after they gave their individual responses.

After the individual phase, participants entered the group phase. Since participants had viewed different stimulus sequences that were produced by the same coin, they engaged in a group discussion to aggregate the individually gathered evidence and produce a real group response. Similar to the individual responses, groups reported a decision and rated their confidence in that decision. We label these responses based on real group discussions \emph{reported group decision} and \emph{reported group confidence} and later compare them to the \emph{simulated group decision} and \emph{simulated group confidence}, which we obtain from statistically aggregating individual responses using CWMV. Groups were allowed to give a group response not earlier tha{\color{black}n} 30 seconds and discussions usually did not last longer than 2 minutes. 

The ideal group responses, $y^*_g$ and $c^*_g$, can be determined by adding the number of red and blue disks from all three stimulus sequences shown to the participants. Then, the same calculations as for ideal individual responses can be applied to compute the ideal group responses. Alternatively and equivalently, aggregating ideal individual responses using CWMV (Equations~\ref{eqn:ycwmv} and \ref{eqn:ccwmv}) also produces the ideal group responses because CWMV aggregates confidences in the theoretically correct way.

Across the 12 experimental trials, there were four Scenarios I -- IV. Each scenario was defined by three stimulus sequences: A, B and C. Table~\ref{tab:scenarios} shows the ideal decision and confidence for each stimulus sequence in each scenario as well as the ideal group responses. Each participant saw one of those sequences from the current scenario. These scenarios were repeated three times in a randomized order for a total of 12 trials and the stimulus sequence that participants saw (A, B and C) were rotated so that participants viewed different stimulus sequences when a scenario was repeated. Importantly, Scenarios II and IV were designed so that MV and CWMV yield different predictions because the most confident individual should—according to CWMV—outweigh the relatively unconfident majority.

At the end of the group phase in each trial, the group was incentivized for giving an accurate group confidence rating. They entered a lottery in which the group could win money depending on how accurate the reported group confidence was. This lottery, the Matching Probability method, was conceived by \citeA{massoni2014}, see also \citeA{dienes2010gambling}. The probability to win in this lottery is maximized if the group confidence is neither under- nor overestimated. Participants were instructed about the rules of this lottery and it was emphasized that chances to win are best when confidence ratings reflect the probability of the group decision to be correct. In each trial, groups could win 0.60 EUR (app{\color{black}roximately} 0.66 USD). Across 12 experimental trials, groups could win a total maximum of 7.20 EUR (7.90 USD) in addition to their compensation for participation. The sum was split equally among the three participants of the group. We did not apply this lottery for individual confidence ratings because these have already been shown to be reliable \cite{griffin1992} so that incentivation was not necessary in the individual phase. In contrast, incentivation was applied in the group phase because we assumed that it is important to additionally motivate participants there and keep them engaged in the group discussions.

\subsection*{Formal cognitive modeling of CWMV}

CWMV is the theoretically optimal way of aggregating individual responses. Real groups on the other hand may deviate from CWMV {\color{black}in various ways}. To measure {\color{black}these} deviations, we introduce four parameters into the CWMV framework in order to capture different aspects in which real groups deviate from CWMV:
\begin{itemize}
    \item $\sigma_i$: \emph{precision of individuals} in recovering the ideal confidence in their reported confidence ratings,
    \item $\beta$: \emph{equality {\color{black}effect}}, or, tendency of groups to weigh individual votes more equal or more extreme than {\color{black}CWMV would} based on the individual confidences,
    \item $\gamma$: \emph{group confidence {\color{black}effect}} determining whether groups tend to over- or underestimate their confidences, and
    \item $\sigma_g$: \emph{precision of groups} {in \color{black}determining the group confidence in accordance with CWMV simulations based on the} individual confidence ratings{\color{black}.} 
\end{itemize}

We estimate individuals' precision, $\sigma_i$, in recovering the true strength of evidence of the displayed stimuli sequences. We assume that individuals are not able to determine the ideal confidence but, instead, their actual responses will scatter around the ideal values. We describe this by an error term $\epsilon_i$:
\begin{linenomath*}
    \begin{equation*}
      c_i = c_i^* + \epsilon_i.
    \end{equation*}
\end{linenomath*}
This error term $\epsilon_i$ is normally distributed with mean zero {\color{black}(reflecting no absolute bias in individual confidence reports in accordance with \citeNP{griffin1992})} and standard deviation~$\sigma_i$. This standard deviation characterizes individuals' precision in recovering the true confidence. An ideal observer would be perfectly precise and make no errors, $\sigma_i = 0$, whereas larger values of $\sigma_i$ indicate less precision. 

Individuals make incorrect decisions if the actual confidence deviates below the 50\% threshold resulting in the complementary confidence towards the incorrect decision.
\begin{linenomath*}
    \begin{align*}
      y_i = \begin{cases}
    y_i^* \text{~~~~(correct),~} & \text{if $c_i \geq 0.5$}.\\
    -y_i^* \text{~~(incorrect),~} & \text{otherwise}.
  \end{cases}
    \end{align*}
\end{linenomath*}
In our experiment, we used a half scale ranging from 50\%--100\% towards the decision made by the participant. For correct estimation, we transform the reported confidences into a full scale ranging from 0\%--100\% towards the correct decision (see \citeNP{olsson2014measuring}) by inverting confidences towards the incorrect decision. For example, when an individual responded incorrectly with a confidence of 60\%, we transform the confidence to $c_i = .4$ (40\% towards the correct alternative) in order to estimate $\epsilon_i$ in each trial and thereupon $\sigma_i$.

Note that confidence ratings cannot be higher than 100\%, which potentially causes a ceiling effect \cite{griffin2004perspectives}. However, in our experiment, ideal confidences for individual responses only range up to a maximum of 88\% (Scenario~III, Individual~A in Table~\ref{tab:scenarios}) so that there is enough room for positive deviations, $\epsilon_i$, to avoid a large ceiling effect here.

Furthermore, we introduce the parameter $\beta$ to estimate the equality effect capturing whether real groups weighted individual responses in a {\color{black}way that deviates from CWMV}. This parameter acts upon the weights $w_i$ as an exponent:
\begin{linenomath*}
    \begin{equation}
      y_g^\text{CWMV}(\beta) = \text{sign}\left(\sum_i w_i^{\beta} y_i\right) \text{.}
      \label{eqn:ycwmvfcm}
    \end{equation}
\end{linenomath*}
As the name suggests, the equality {\color{black}effect} models groups assigning more equalized weights than naive CWMV, which is conceptually similar to the approach by \citeA{mahmoodi2015equality} but our model is technically different because we incorporate it into the CWMV framework. Here, the equality {\color{black}effect} can vary between zero and infinity, $\beta \in [0; \infty]$. In the edge case of $\beta=0$, every weight would be transformed equally to $w_i^0 = 1$ producing the special case of (unweighted) MV. On the other hand, $\beta=1$ would leave weights unchanged, $w_i^1 = w_i$, and would produce undistorted CWMV. Values in between, $0 < \beta < 1$, would represent some compromise in which individual confidences are considered to some extent but groups {\color{black}tend to equalize} those weights. On the other side of the spectrum, larger values of $\beta>1$ would represent an exaggeration of differences between weights so that the most confident individual's vote has a disproportionately large impact. In such situations, the most confident individual would tend to decide the vote single-handedly, which is equivalent to the predictions from maximum-confidence slating \cite{koriat2012two}.

We additionally estimate whether groups under- or overestimate their group confidence, which is captured in the group confidence {\color{black}effect} $\gamma$:
\begin{linenomath*}
    \begin{equation}
      c_g^\text{CWMV}(\beta, \gamma) = \frac{1}{1+\exp({-\gamma|\sum_i w_i^\beta y_i|})}  ~\text{.}
      \label{eqn:ccwmvfcm}
    \end{equation}
\end{linenomath*}
The group confidence {\color{black}effect} allows for a non-linear scaling of the group confidences. This parameter can also vary between zero and infinity, $\gamma\in[0;\infty]$, where $\gamma<1$ represents groups underestimating their confidence relative to the ideal statistical aggregation of individual responses, whereas $\gamma>1$ represents an overestimation of group confidences. The special case of $\gamma = 1$ recovers undistorted (naive) CWMV.

Note that the equality {\color{black}effect} $\beta$ modifies individual weights and can potentially change the simulated group decision. In contrast, the group confidence {\color{black}effect} $\gamma$ only modifies a group's final confidence (hence, it does not appear in Equation~\ref{eqn:ycwmvfcm} were the simulated group decision is determined). {\color{black}These two parameters capture deviations from naive CWMV simulations in a descriptive manner. For cautionary accounts against normative interpretations, see \citeA{gigerenzer2018bias,le2011rational}; and \citeA{neth2016rational}.}

Finally, we introduce an error term to the group confidence $c_g = c_g^\text{CWMV}(\beta, \gamma) + \epsilon_g$. This error term $\epsilon_g$ acts similar to the error term of individual confidence ratings. It is normally distributed with mean zero and standard deviation $\sigma_g$, where smaller values indicate higher precision of the group discussion process matching the ideal aggregation. 

For estimation, the individual precision $\sigma_i$ was measured by computing the average of sample variances across individuals and taking the square root. For the group parameters, we performed a grid search in which we varied $\beta$ and $\gamma$ in $[0, 2]$ and $\sigma_g$ in $[0, 0.3]$ (larger values produced worse fits) with step sizes of $0.01$. For each group, we chose the parameter combination that produced the maximum likelihood for the observed data using Equations~\ref{eqn:ycwmvfcm} and \ref{eqn:ccwmvfcm} to predict the real group responses. 

We validated this approach by conducting multiple parameter recovery simulations as suggested by \citeA{wilson2019ten}: We simulated data based on our model for fixed values of $\sigma_i$, $\beta$, $\gamma$ and $\sigma_g$ and demonstrated that our estimation approach recovered the ground truth parameters, see open material for details.

\section*{Results}

\begin{figure}[!ht]
	\centering
    \includegraphics[page=1,width=.35\textwidth]{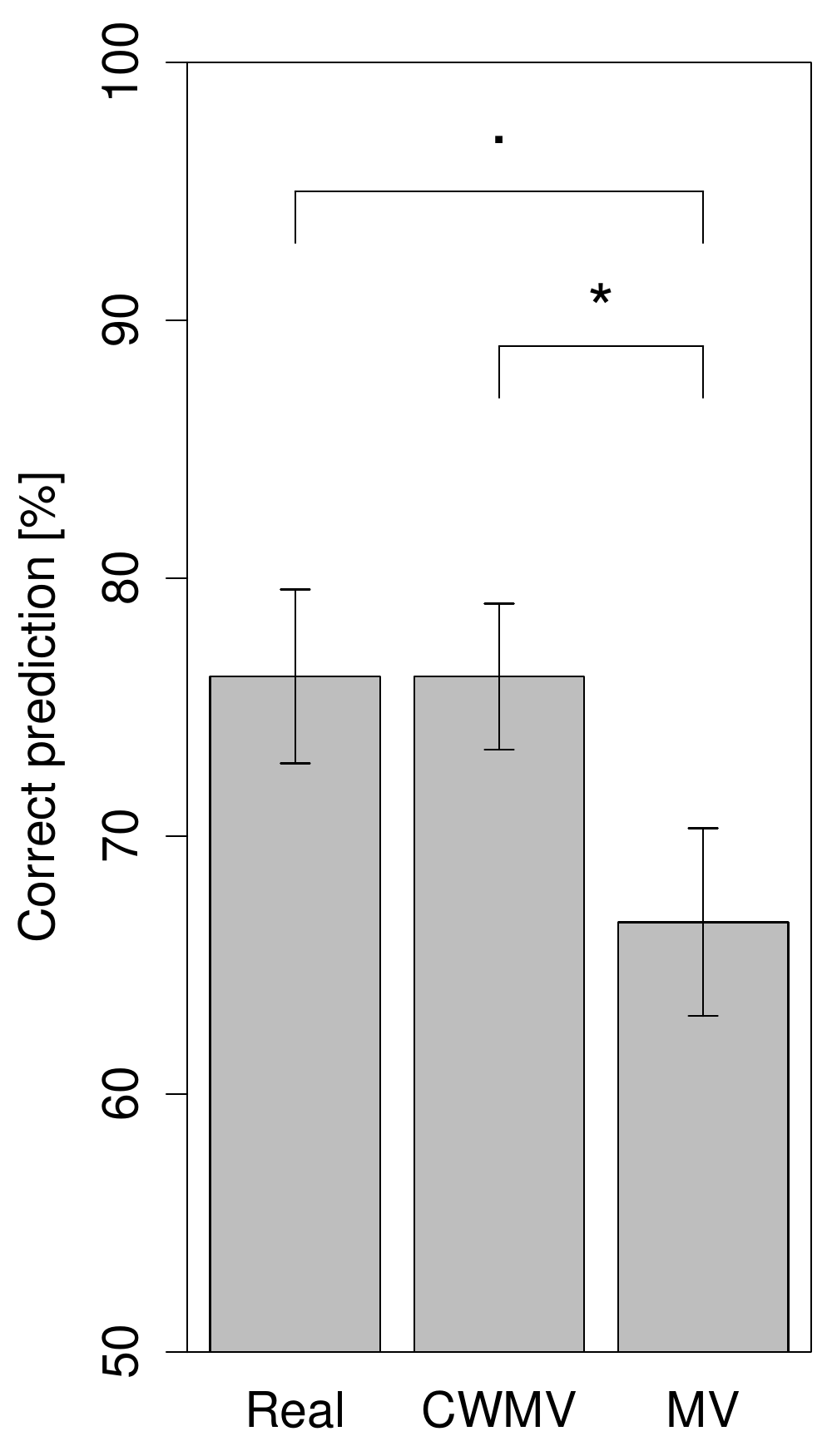}
    \caption{\textbf{Performance of real versus simulated groups.} Comparing the accuracy of real group decisions to simulated group decisions using either CWMV or MV for aggregation of the individual decisions and confidences. Simulated groups based on CWMV predict the performance of real groups very well, while simulated groups based on MV underestimate the performance of real groups. Error bars indicate standard errors of the mean computed across groups. \newline $^*p<.05$. $^{\boldsymbol{\cdot}}p<.1$.}
      \label{fig:accuracyBarplot}
\end{figure}

We compared the average and median performance of real versus simulated groups, see Figure~\ref{fig:accuracyBarplot} and see Additional File~1 for estimates of each group. Real groups reported the correct (ideal) decision in $76.2\%$ ($SEM=3.4\%$) of the trials (Mdn~=~$75.0\%$, IQR~=~$75.0$--$83.3$). CWMV adequately simulated the average performance of real groups with $76.2\%$ ($SEM=2.8\%$, Mdn~=~$75.0\%$, IQR~=~$70.8$--$83.3$). In contrast, simulating group decisions using unweighted MV produced a lower accuracy of $66.7\%$ ($SEM=3.6\%$, Mdn~=~$66.7\%$, IQR~=~$62.5$--$75.0$) compared to CWMV {\color{black}with a mean difference of $M=9.5\%$ ($SEM = 3.4\%$)}, $t(6) = 2.83$, $p = .030$. Comparing MV to real groups yielded a trend towards the same difference, {\color{black}$M=9.5\%$ ($SEM = 4.6\%$)}, $t(6) = 2.07$, $p = .084$. We conducted two-sided, exact binomial tests to confirm this pattern: MV simulations were less accurate than CWMV simulations ($p=.016$) and real group decisions ($p=.016$). 

\subsection*{Real versus ideal responses}

\begin{figure*}[!ht]
    \includegraphics[page=1,width=1\textwidth]{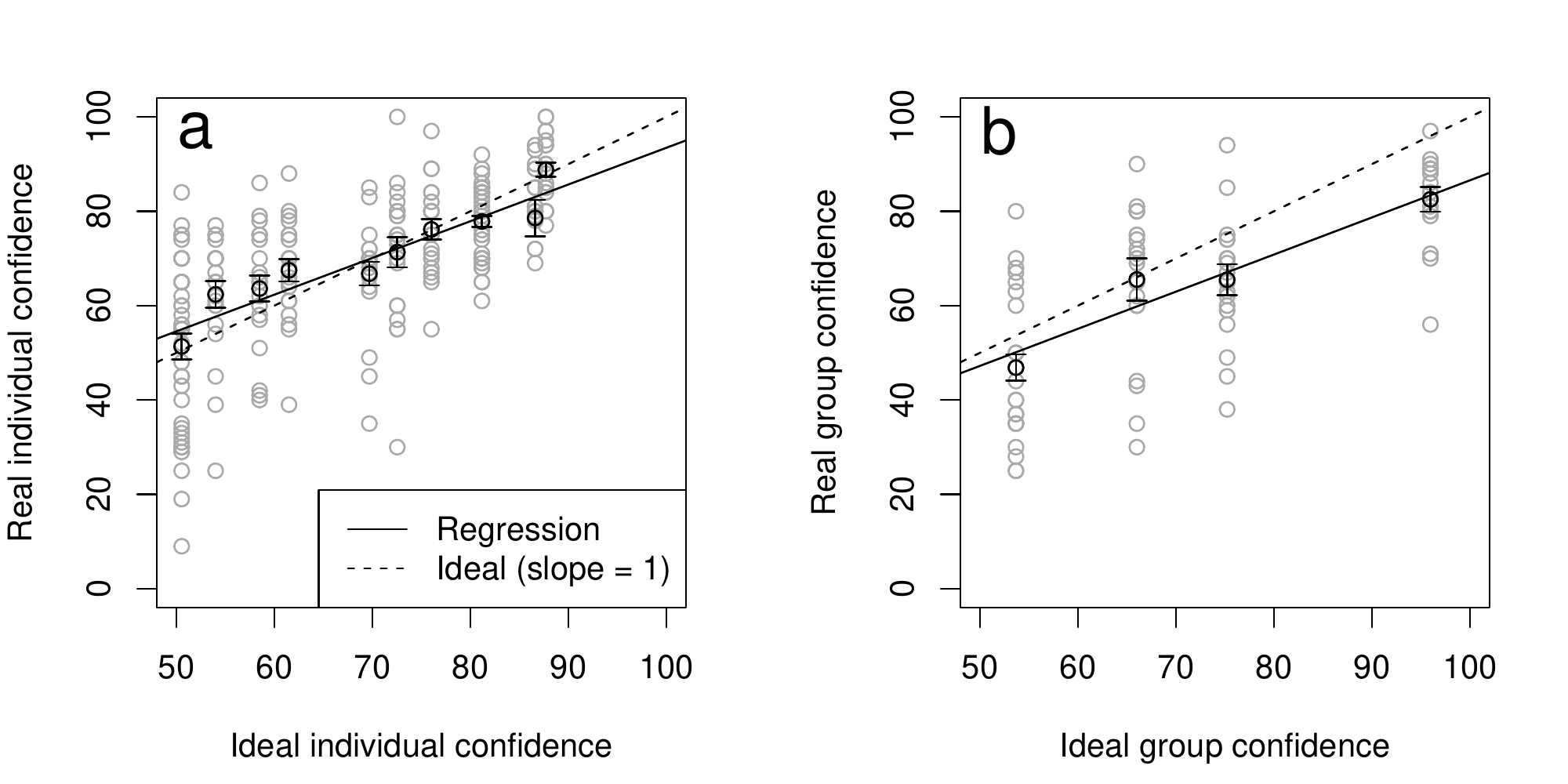}
    \caption{\textbf{Comparing ideal versus reported confidences from individuals and groups.} Ideal confidence (x-axis) range from 50\% to 100\% in accordance to Table~\ref{tab:scenarios}. In contrast, reported confidences (y-axis) range from 0\% to 100\% because we flipped confidence ratings in cases where an incorrect decision was given (e.g., a reported confidence of 60\% towards the incorrect decision is displayed as a confidence of 40\% here). In (a), reported confidences from individuals (y-axis) are compared to the ideal values (x-axis; cf. $c_1^*$, $c_2^*$, and $c_3^*$ from Table~\ref{tab:scenarios}). Similarly in (b), reported confidences from groups (y-axis) are compared to the ideal values (x-axis; cf. $c_g^*$ from Table~\ref{tab:scenarios}). Black points indicate mean values—averaged across individuals in (a) and across groups in (b)—for each ideal value. Grey points indicate single trial responses. Error bars indicate standard errors of the mean.}
      \label{fig:reportedVsIdeal}
\end{figure*}

Individual confidence ratings were well aligned with the ideal confidences, see Figure~\ref{fig:reportedVsIdeal}a. The average correlation between reported versus ideal confidences across individuals was $\bar{r}=.73$, $95\%$~CI~$[0.64,0.80]$ (we used Fisher's $z$-transformation for combining correlations into averages). This finding replicates \citeA{griffin1992} showing that individual participants are able to evaluate the ambiguity in the presented stimulus sequences and report their confidences in form of subjective probabilities. Estimating the precision of individuals, we observed that reported confidences scattered around ideal confidences with a standard deviation of $\sigma_i = 13.3\%$, $SD = 6.6$, $95\%$~CI~$[9.8, 16]$.

However, confidence reports showed {\color{black}systematic deviations}. In hard (difficult) trials with low ideal confidences, individuals overestimated those confidences. This is reflected in regression lines on average being at $M = 55\%$, $95\%$~CI~$[50.2, 58.8]$, where they should be at 50\%. Additionally, high confidences were underestimated. The average slope of regression lines was lower than the ideal value 1, $\bar{b}=0.78$, $95\%~$CI~$[0.61, 0.95]$. A slope of 1 would have indicated that ideal and reported confidences increased equally, whereas, here, the observed slope below 1 indicated that increasing the true evidence strength from the presented stimulus sequences only led to a diminished increase in confidence.

Group confidence ratings showed a somewhat similar pattern, see Figure~\ref{fig:reportedVsIdeal}b (we again present median values). The average correlation between reported and ideal group confidences was high, $\bar{r}=.71$, $95\%$~CI~$[.57,.80]$, Mdn~=~$.71$, IQR~=~$.64$--$.79$, {\color{black}but there was a relatively large root mean squared error, $RMSE = 0.16$}. Real groups did not deviate from ideal values at low confidences: The regression lines at the ideal 50\% were $M=47\%$, $95\%$~CI~$[40.7, 53.7]$, Mdn~=~$48.6\%$, IQR~=~$42.5$--$50.6$. Nevertheless, groups (similar to individual participants) underestimated high confidences resulting in an attenuated average slope relative to the ideal value of 1, $\bar{b} = 0.79$, $95\%$~CI~$[0.58, 0.99]$, Mdn~=~$0.77$, IQR~=~$0.73$--$0.96$. {\color{black}The large RMSE reflects this divergence for high confidences.} Exact binomial tests confirmed these results: All groups had a correlation above $0$ and a slope below $1$, both $p=.016$, but intercepts scattered around 50\%, $p=.336$. Note that we avoid common problems of regression in the context of over- versus underconfidence estimation since our regressions use the fixed ideal confidences as independent variables (x-axes in Figure~\ref{fig:reportedVsIdeal}), which exhibit no estimation error that would otherwise have lead to a biased analysis \cite{fiedler2012more,olsson2014measuring}.

\subsection*{Real versus simulated group responses}

\begin{figure*}[!ht]
    \includegraphics[page=1,width=1\textwidth]{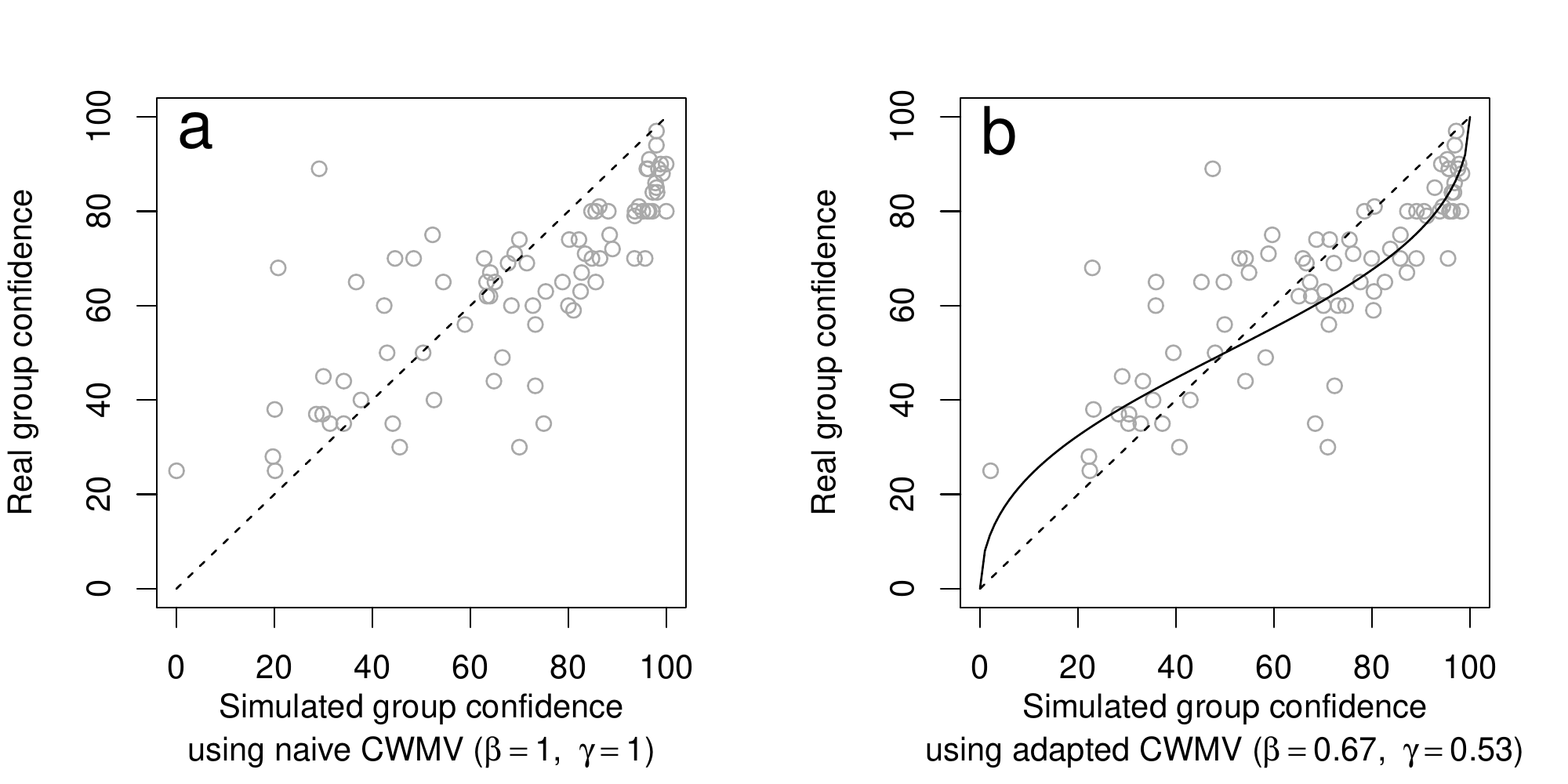}
    \caption{\textbf{Comparing real versus simulated group responses from statistically aggregating individual responses.} We used individual responses to simulate group confidences via CWMV (x-axis). These simulations predict responses from real, interacting groups (y-axis). In (a), we used naive CWMV as in Equations~\ref{eqn:ycwmv} and \ref{eqn:ccwmv}. The dashed line represents predictions from naive CWMV. This is equivalent to our formal cognitive modeling with equality {\color{black}effect} $\beta = 1$ and group confidence {\color{black}effect} $\gamma = 1$. In (b), we estimated the equality {\color{black}effect}, $\beta = 0.67$, and group confidence {\color{black}effect}, $\gamma = 0.53$, see Equations~\ref{eqn:ycwmvfcm} and \ref{eqn:ccwmvfcm}. This model predicts real group responses (solid line) but incorporates the fact that real groups treated individual votes more equal and displayed an underconfidence {\color{black}effect}. In both subfigures, confidence ratings are inverted for incorrect responses. For example, the point (34\%, 35\%) in (a) corresponds to a trial with a simulated confidence of 66\% and a reported confidence of 65\% with both decisions being the same but incorrect, hence, both confidences were inverted.}
      \label{fig:reportedVsSimulated}
\end{figure*}

Responses from real, interacting groups were well predicted by simulated responses using CWMV. Naive CWMV (Equations~\ref{eqn:ycwmv} and \ref{eqn:ccwmv}) produced an average correlation between reported and simulated confidences of $\bar{r} = .83$, $95\%$~CI~$[0.56,0.94]$, Mdn~=~$.82$, IQR~=~$.64$--$.92$. {\color{black}Despite the high correlation, there was still a large discrepancy, $RMSE = 0.17$, reflecting {\color{black}deviations of real responses from naive CWMV}}, see Figure \ref{fig:reportedVsSimulated}a.

We applied our formal cognitive model to estimate in how far real groups deviated from naive CWMV. The equality {\color{black}effect} $\beta$ was on average $M = 0.67$, $SD = 0.30$, $95\%$~CI~[0.38, 0.95], Mdn~=~$0.74$, IQR~=~$0.38$--$0.95$. This indicates that groups used confidences similar to CWMV but {\color{black}tended} towards equalizing those weights. Votes from confident individuals were given more impact on the group decision compared to unconfident individuals but not to the {\color{black}extent suggested by CWMV}. We observed trials in which the most confident individual is overruled by the majority and the tipping point at which this happened was earlier than what naive CWMV simulations predict. This observation is captured in the equality {\color{black}effect} estimate being smaller than 1, $\beta = 0.67~<1$. 

It is noteworthy that, since $\beta$ estimates are always larger than zero, it is a priori expected to obtain an above zero average simply due to random errors. To account for this, we performed a randomization test where we randomly permutated individual confidences and estimated $\beta$ from the resulting data set. Since the confidences in these randomized data sets are not indicative of the group's decision, the true equality {\color{black}effect} is zero here. From 1,000 of such randomizations, we found that 95\% of the obtained $\beta$ estimates were below $0.4$. This confirms that groups in our experiment (with $\beta = 0.67$) did take confidences into account ($\beta>0$) but only to an attenuated extent ($\beta<1$). 

The group confidence {\color{black}effect} $\gamma$ was on average $M = 0.53$, $SD = 0.09$, $95\%$~CI~$[0.45, 0.61]$, Mdn~=~0.62, IQR~=~$0.55$--$0.74$, indicating that real groups tend to underestimate ($\gamma < 1$) {\color{black}their confidence compared to CWMV simulations based on the} individual responses. In Figure~\ref{fig:reportedVsSimulated}b, this underestimation {\color{black}effect} corresponds to a predicted curve (solid line) below the ideal values (dashed line).

The average group precision was $\sigma_g = 11\%$ (root mean square; $SD = 4$) with Mdn~=~10\%, IQR~=~7\%--12\%. This precision of group confidences is comparable to the precision of individual confidences.

{\color{black}The adapted CWMV model using $\beta = 0.67$ and $\gamma = 0.53$ predicted confidences that are correlated with reported confidences to the same degree as naive CWMV, $\bar{r} = .84$, $95\%$~CI~$[0.68,0.93]$, Mdn~=~$.84$, IQR~=~$.72$--$.92$. But in absolute terms, this adapted CWMV model matched the reported confidences better ($RMSE = 11\%$) than naive CWMV ($RMSE = 17\%$, mentioned above), $t(6) = 5.24$, $p < .002$, because the adapted model simulates group responses with an equality and underconfidence {\color{black}effect}.}

Note that going from Figure \ref{fig:reportedVsSimulated}a to Figure \ref{fig:reportedVsSimulated}b, points are shifted along the x-axis because the equality {\color{black}effect} $\beta = 0.67$ changes simulated confidences and can even change the simulated decision (points crossing the $50\%$ border in the $x$ direction). The extent of these shifts depend{\color{black}s} on the exact constellation of individual confidences. On the other hand, the group confidence {\color{black}effect} $\gamma = 0.53$ only maps the resulting, simulated confidences in a non-linear way to the reported confidences (solid curved line). This parameter reflects that groups were less confident in their decisions th{\color{black}a}n what naive CWMV predicted. 

\subsection*{Model comparison of group response simulations}

To evaluate our adapted CWMV model, we compared the full model to three special case models in which we fixed one parameter at a time (first $\gamma = 1$, second $\beta = 0$ and third $\beta = 1$). For this comparison, we computed the Bayesian Information Criterion (BIC; \citeNP{schwarz1978estimating}). Using the Akaike Information Criterion (\citeNP{akaike1973information}) instead of BIC yielded qualitatively identical results. Smaller BIC values indicate a better fit relative to the number of parameters in the model. For the full model, the total score (sum across groups) was $BIC_\text{full} = -101$. 

As a first comparison, we pitch the full model against a model that fixes the group confidence {\color{black}effect} $\gamma = 1$ but keeps the equality {\color{black}effect} $\beta$ free. This model assumes that groups may only deviate from naive CWMV in terms of how they assign weights to the individual votes but exhibit no general over- or underconfidence. Here, the total score was $BIC_{\gamma=1} = -59$ indicating a worse fit as compared to the full model. The Bayes factor resulting from the BIC scores of the two models (e.g., see \citeNP[chapter 11]{farrell2018computational}) clearly supported the full model, $BF_{\text{full}/\gamma=1}~>~1000$. This supports the notion that {\color{black}group responses are best characterized by an overall underconfidence effect}.

The second comparison fixes $\beta = 0$ but keeps $\gamma$ free. This model is equivalent to an (unweighted) MV with group confidence {\color{black}effect}. Here, the total score was $BIC_{\beta=0} = -63$, again supporting the full model, $BF_{\text{full}/\beta=0} > 1000$. This indicates that participants incorporate confidence ratings in the group discussion.

For the third comparison, we fix $\beta = 1$: This model assumes that real groups weigh individual votes exactly according to undistorted CWMV but still allows for an overall confidence {\color{black}effect} of the group since $\gamma$ is free. This model was on par with the full model, $BIC_{\beta=1} = -102$, with an inconclusive Bayes factor, $BF_{\text{full}/\beta=1} = 0.71$. This indicates that, according to the BIC criterion, fixing $\beta = 1$ {\color{black}did not perform worse (when accounting for the additional free parameter) than the full model, which keeps $\beta$ free. On the other hand, when performing a model fit comparison irrespective of the number of parameters \cite[chapter 10]{farrell2018computational}, the full model performs better than that with fixed $\beta = 1$, $\chi^2(7) = 16.9$, $p = .018$. To confirm that incorporating the equality {\color{black}effect} $\beta$ as a free parameter in our model  conveys an advantage even when weighing parsimony against model fit, future research with increased sample sizes are necessary.}

\section*{Discussion}

Including confidence ratings in the theoretically optimal way using CWMV increases the simulated group performance over MV. Real groups are more accurately represented by CWMV when individuals provide reliable and independent confidence ratings. Even though real groups consider confidence ratings {\color{black}similar to CWMV}, they tend to treat individual responses more equally giving more confident individuals less impact on the group decision than naive CWMV simulations, which is {\color{black}consistent with an} equality bias \cite{bang2017,mahmoodi2015equality}. Additionally, groups tend to underestimate their confidences.

In our study, individuals were overconfident in hard (difficult) trials and underconfident in easy trials—a finding often referred to as the hard-easy effect \cite{gigerenzer1991probabilistic}. Hard trials allow participants to make a correct decision about 50\% of the time but reported confidences were larger. In contrast, easy trials allow for close to 100\% confidences but reported confidences were strictly lower. This hard-easy effect, or underextremity \cite{griffin2004perspectives}, can be explained by a regressive tendency \cite{moore2008trouble}. That is, participants were biased towards their prior belief to observe trials with moderate difficulty. But it can also be explained by a bias introduced through the response format: \citeA{olsson2014measuring} argue that a half scale (50\%--100\%), as it is often used, biases participants to respond closer to the center of the scale.

In contrast, real groups did not tend to be overconfident for hard trials in our setting but real groups exhibited overall underconfidence in a double sense: First, group confidences were lower than ideal responses (see Figure~\ref{fig:reportedVsIdeal}b). Second, group confidences were lower than {\color{black}determined by CWMV simulations based on} individual responses (see Figure~\ref{fig:reportedVsSimulated}b). 

Interestingly, confidence ratings reflected subjective probabilities rather than consistency in our study. For example, we presented a stimulus sequence that is suited to evoke a low ideal confidence of 54\% (see Table 1, Scenario III, Individual B). For this sequence, participants gave the correct decision in 85.7\% of the trials and reported confidences relatively close to the ideal confidence with an average of 62\% (see Figure~\ref{fig:reportedVsSimulated}, second black point from left). In other words: Participants consistently determined the correct decision but nevertheless reported in their confidence ratings that the strength of evidence was quite low as intended.

 One limitation that our well controlled setting cannot account for are situations in which individuals consensually reach incorrect decisions with high confidence (see \citeNP{koriat2015two,koriat2017can,litvinova2019experts}). In such situations, confidences towards the incorrect decision are aggregated and can lead to high group confidences towards incorrect decisions. In how far CWMV can adequately reflect real groups in these situations remains to be shown because consensually incorrect decisions were too rare in our setting to allow inferences, see bottom left quadrants in Figure~\ref{fig:reportedVsSimulated}.

Further insight into group processes can be gained by fixing the ideal group confidence and varying the constellation of individual confidences. For example, our Scenario II determined an ideal group confidence of 75\% based on one confident individual (76\% for biased coin) and two almost uninformative individuals (51\% for fair coin). The same ideal confidence of 75\% would come from three equally confident members (59\% for biased coin). CWMV predicts the same ideal confidence but real groups may behave differently in these two cases. From our current estimates of the equality bias ($\beta = 0.67$), we predict that real groups are more confident in the latter constellation where each individual contributes an equal confidence as compared to a situation where only one individual is very confident.

Our controlled setting provided optimal conditions for CWMV with independent confidence ratings but it was rather artificial. This allowed us to verify that groups are indeed able to perform confidence weighting to some extent. However, in real world tasks, bad calibration of confidences may prevent simulated groups to perform as well as real groups. For example, \citeA{klein2015} observed that individuals could not report well calibrated confidence ratings but real groups still outperformed simulations using MV. One possibility is that individuals failed to rate their confidence in a comparable way when verbal scales were used instead of numeric scales \cite{windschitl1996measuring}: Klein and Epley used a 9-point Likert scale from "not at all confident" (1) to "very confident" (9). Nevertheless, participants might have been able to share calibrated confidences in the real group discussions. This could have led to a better performance of real compared to simulated groups.

Another possible reason for real groups outperforming simulated groups is that the assumption of independence is violated. These—arguably more realistic—situations have been investigated under the name of \emph{hidden profiles}, where a hidden profile determines the distribution of information that is either common among or unique to individuals \cite{stasser2003hidden,stasser2020collective}. Distinguishing between evidence that is held by all individuals of a group versus evidence that is uniquely known by few individuals is a crucial aspect of successful real groups \cite{mercier2016argumentative}. Consider again the example of three individuals deciding whether a suspect is guilty or not. Say, individuals have in total five pieces of evidence: two incriminating, $I_1$ and $I_2$, and three exonerating, $E_1$, $E_2$ and $E_3$. All individuals know all the incriminating evidence but each individual knows only one unique piece of exonerating evidence. That is, the first individual knows $I_1$, $I_2$, and $E_1$; the second knows $I_1$, $I_2$, and $E_2$; and the third knows $I_1$, $I_2$, and $E_3$. For each individual there is more incriminating evidence and each would decide 'guilty' with some confidence. Incorrectly assuming independence, CWMV would simulate the group decision to be guilty as well. However, a real group might lay out all the evidence, find in total more exonerating evidence, and decide 'not guilty'. 

There are some approaches to handle such dependencies formally \cite{kaniovski2011optimal,shapley1984optimizing,stasser1987effects} each coming with its own set of particular, additional assumptions. To sketch the approach that we find most promising: CWMV could be applied not to the potentially dependent individual responses but to the independent pieces of evidence, with confidences indicating the strength of each piece of evidence. Incorporating CWMV in this way could improve theoretical predictions: Rather than comparing group performance to the best individual (as is often done), CWMV-inspired approaches may provide a more adequate baseline for group performance even when information is distributed in a way that violates the independence assumptions for individual responses. 

\section*{Conclusion}

Confidence ratings of individuals play an important role in real group decisions and can be used to increase simulated group performance. In a controlled setting, real groups have proven to aggregate confidences in a way that is to some extent consistent with the CWMV even though they tend to treat individual responses more equal {\color{black}and with lower confidence than when using CWMV simulations}. Developing group simulation methods (for example to account for dependencies) and comparing simulated group decisions using those methods to real group decisions will deepen our understanding of real world group discussions.

%%%%%%%%%%%%%%%%%%%%%%%%%%%%%%%%%%%%%%%%%%%%%%
%%                                          %%
%% Backmatter begins here                   %%
%%                                          %%
%%%%%%%%%%%%%%%%%%%%%%%%%%%%%%%%%%%%%%%%%%%%%%

\section*{Abbrevations}

MV: Majority vote; CWMV: Confidence weighted majority vote

\section*{Ethics approval and consent to participate}

Written informed consent was obtained from all
participants. The experiment was conducted in accordance with the 1964
Declaration of Helsinki and with the ethical guidelines of the German
Psychological Society (DGPs) and the Professional Association of
German Psychologists (BDP) (2005, C.III). 

\section*{Consent for publication}
Not applicable.

\section*{Availability of data and material}
The experimental code and data set supporting the conclusions of this article is available in the Open Science Framework repository (doi: \url{https://doi.org/10.17605/OSF.IO/G69KZ}), \url{https://osf.io/g69kz/?view_only=5c068a3d93cd4915ac6e86c53b19ccdc}.

\section*{Competing interests}
The authors declare no competing interests.

\section*{Funding}
This project is supported by the Deutsche Forschungsgemeinschaft (DFG,
German Research Foundation) through the CRC 1233 ``Robust Vision'',
project number 276693517; 
the Institutional Strategy of the University
of T{\"u}bingen (DFG, ZUK 63); and the Cluster of Excellence “Machine
Learning: New Perspectives for Science”, EXC 2064/1, project number
390727645. 

\section*{Author's contributions}
S.M., D.M.B.S., and V.F. conceptualized the experiment and the analyses. D.M.B.S. and S.M. performed the experiment and implemented the analyses. S.M., V.F., D.M.B.S., and U.L. wrote manuscript.  

\section*{Acknowledgements}
We would like to thank Frieder G{\"o}ppert, Catarina Amado and Iris A. Zerweck for their helpful feedback. \newline

\section*{Author details}
$^1$ Experimental Cognitive Science, Department of Computer Science, University of T{\"u}bingen. $^2$ Theory of Machine Learning, Department of Computer Science, University of T{\"u}bingen. $^3$ Max Planck Institute for Intelligent Systems, Tübingen, Germany

\bibliographystyle{apacite}

%DONOTRUNBIBTEX

~\newpage 
\onecolumn
\appendix
\section{~}~

\begin{table*}[!ht]

\caption{\fontseries{m}\selectfont \textbf{Summary statistics and parameter estimates.} For each group, we report the average accuracy (\%-correct responses across 12 trials) for real and simulated groups, regression coefficients $a$ and $b$ ($\text{real confidence} = a + b\times\text{ideal confidence} + \epsilon$), and fitted parameters from our formal cognitive modeling approach: equality bias ${\beta}$, group confidence bias ${\gamma}$ and group precision $\sigma_g$. Additionally, we report mean, standard deviation as well as the more robust median and quartiles.}
\label{tab:parameterEstimates}
\centering
\begin{tabular}{ccccccccc}                                                        
  \toprule                                                                        
 Group & \multicolumn{3}{c}{Correct Predictions [\%]} &  \multicolumn{2}{c}{Regression} &  \multicolumn{3}{c}{Formal Cognitive Model} \\                            
 \cmidrule(lr){2-4} \cmidrule(lr){5-6} \cmidrule(lr){7-9} &  Real &  CWMV &  MV & 
 Intercept $a$ &  Slope $b$ &  ${\beta}$ &  ${\gamma}$ &  ${\sigma}_g$ \\         
 \midrule                                                                         
  1 & 75.0 & 75.0 & 66.7 & 43.5 & 0.94 & 0.68 & 0.62 & 6.0 \\                     
  2 & 75.0 & 83.3 & 75.0 & 49.5 & 0.70 & 1.23 & 0.51 & 8.0 \\                    
  3 & 58.3 & 83.3 & 58.3 & 37.4 & 0.77 & 0.53 & 0.38 & 10.0 \\                   
  4 & 83.3 & 66.7 & 66.7 & 58.4 & 0.35 & 0.80 & 0.52 & 11.0 \\                   
  5 & 75.0 & 75.0 & 75.0 & 41.4 & 0.99 & 0.23 & 0.46 & 13.0 \\                   
  6 & 83.3 & 83.3 & 75.0 & 51.8 & 1.00 & 0.62 & 0.61 & 7.0 \\                    
  7 & 83.3 & 66.7 & 50.0 & 48.6 & 0.76 & 0.57 & 0.59 & 17.0 \\                   
   \midrule                                                                       
  Mean   & 76.2 & 76.2 & 66.7 & 47.2 & 0.79 & 0.67 & 0.53 & 10.9 \\                  
  SD     & 8.9  & 7.5  & 9.6  & 7.0  & 0.23 & 0.30 & 0.09 & 3.8 \\                       
  Median & 75.0 & 75.0 & 66.7 & 48.6 & 0.77 & 0.62 & 0.52 & 10.0 \\              
  25\%-Quantile & 75.0 & 70.8 & 62.5 & 42.5 & 0.73 & 0.55 & 0.48 & 7.5 \\        
  75\%-Quantile & 83.3 & 83.3 & 75.0 & 50.6 & 0.96 & 0.74 & 0.60 & 12.0 \\       
   \bottomrule                                                                    
\end{tabular}

\end{table*}     

\end{document}